\documentclass[12pt,showpacs]{article}
\usepackage{amsfonts,pifont}
\usepackage{amsmath,chemarrow}
\usepackage{amssymb}
\usepackage{amstext}
\usepackage{amsfonts}
\usepackage{graphicx}
\usepackage{indentfirst}
\usepackage{hyperref,comment}
\makeatletter
\def\ExtendSymbol#1#2#3#4#5{\ext@arrow 0099{\arrowfill@#1#2#3}{#4}{#5}}
\def\RightExtendSymbol#1#2#3#4#5{\ext@arrow 0359{\arrowfill@#1#2#3}{#4}{#5}}
\def\LeftExtendSymbol#1#2#3#4#5{\ext@arrow 6095{\arrowfill@#1#2#3}{#4}{#5}}
\makeatother

\begin{document}
\baselineskip 20pt

\title{Molecular states with hidden charm and strange in QCD Sum Rules}

\author{Cong-Feng Qiao$^{1,2}$\footnote{qiaocf@ucas.ac.cn}\; and\ Liang Tang$^1$\footnote{tangl@ucas.ac.cn}\\[0.5cm]
$^{1}$School of Physics, University of Chinese Academy of Sciences
\\
YuQuan Road 19A, Beijing 100049, China\\[0.2cm]
$^{2}$
Collaborative Innovation Center for Particles and Interaction\\
USTC, HeFei 230026, China\\[0.2cm] }

\date{}
\maketitle

\begin{abstract}

This work uses the QCD Sum Rules to study the masses of the $D_s \bar{D}_s^*$  and $D_s^* \bar{D}_s^*$ molecular states with quantum numbers $J^{PC} = 1^{+-}$. Interpolating currents with definite C-parity are employed, and the contributions up to dimension eight in the Operator Product Expansion (OPE) are taken into account. The results indicate
that two hidden strange charmonium-like states may exist in the energy ranges of $3.83 \sim 4.13 $ GeV and $4.22 \sim 4.54 $ GeV, respectively. The hidden strange charmonium-like states predicted in this work may be accessible in future experiments, e.g. BESIII, BelleII and SuperB. Possible decay modes, which may be useful in further research, are predicted.

\end{abstract}

\section{Introduction}

Recently, the BESIII Collaboration claimed that four charged charmonium-like resonances \cite{Ablikim:2013xfr} were observed in the invariance mass spectra of $(D \bar{D}^*)^{\pm}$, $J/\psi \pi^{\pm}$, $h_c \pi^{\pm}$ and $(D^* \bar{D}^*)^{\pm}$, which were respectively referred as $Z_c(3885)$, $Z_c(3900)$, $Z_c(4020)$ and $Z_c(4025)$. Among them, the $Z_c(3900)$ has been confirmed by the Belle Collaboration \cite{Liu:2013dau} and the CLEO-c experiment \cite{Xiao:2013iha}. So far it is still early to tell whether  $Z_c(3885)$ and $Z_c(3900)$, as well $Z_c(4020)$ and $Z_c(4025)$, are the same state or not. Moreover, in very recently, the $Z_c(4430)$, which was first observed by Belle Collaboration \cite{Choi:2007wga}, is confirmed in LHCb experiment with about 13.9 $\sigma$ of significance \cite{Aaij:2014jqa}. The charmonium-like resonances are of particular interest as they facilitate not only investigating the dynamics of interaction between light quarks and heavy quarks, but also the testing of the standard model itself. Until now, their inner structures have
not been well determined by theory. In the literature possible interpretations include molecular states \cite{Cui:2013yva, Zhang:2013aoa, Dias:2013xfa, Ke:2013gia, Cui:2013vfa, Chen:2013omd}, tetraquark states \cite{Faccini:2013lda, Qiao:2013raa, Qiao:2013dda, Drenska:2009cd, Maiani:2014aja}, and enhancement structures \cite{Chen:2011xk, Chen:2013coa}, etc.

In Refs.\cite{Drenska:2009cd, Maiani:2014aja}, $Z_c(4430)$ was treated as a pure tetraquark state, since there is no open charm meson thresholds close to it. It is tempting to think this state to be a radial excitation of the $Z_c(3900)$, which was analyzed in the framework of QCD Sum Rules \cite{Wang:2014vha}. Of the hidden strange charmonium-like states, the CDF Collaboration observed a narrow structure $Y(4140)$ near the $J/\psi \, \phi$ threshold in the exclusive $B^+ \to J/\psi \, \phi \, K^+$ decay mode produced in the $\bar{p} p$ collision at $\sqrt{s} = 1.96 \, \text{TeV}$ \cite{Aaltonen:2009tz, Aaltonen:2011at}. A scalar molecular state constructed by $D_s^* \bar{D}_s^*$ with $J^{PC} = 0^{++}$ was proposed to interpret the $Y(4140)$ \cite{Wang:2009ue, Albuquerque:2009ak, Zhang:2009st} soon after its observation, and the extracted mass coincided with the experimental data.

Utilizing the QCD Sum Rules, people have successfully estimated the mass of $Z_c(3900)$ and $Z_c(4025)$ with $D \bar{D}^*$ and $D^* \bar{D}^*$ currents respectively \cite{Cui:2013yva, Chen:2013omd}. These successes intrigue enormous interests of theorists. It is of great interest to note that these states indicate that a new class of hadrons has been observed, but without the strange flavor until now. Namely, two resonances are expected to exist near the thresholds of $D_s \bar{D}_s^*$ and $D_s^* \bar{D}_s^*$ with the quantum numbers $J^{PC} = 1^{+-}$. In this work, we calculate their mass spectra with QCD Sum Rules, discuss the implications of our results and suggest their possible decay channels in the summary. We hope this work may be helpful to experiment in exploring the hidden strange charmonium-like states.

Primary formulae are presented after the introduction. In Sec.III, the numerical results and related figures are shown. The last section is a short summary.

\section{Formalism}

This work considers $D_s \bar{D}_s^{*}$ and $D_s^* \bar{D}_s^*$ molecular states with $J^{PC} = 1^{+-}$ via QCD Sum Rules \cite{Shifman, Reinders:1984sr, Narison:1989aq, P.Col}. The calculations of the QCD Sum Rules are based on the correlator constructed by two hadronic currents. For an axial vector state, the two-point correlation function is given by:
\begin{eqnarray}
\Pi_{\mu \nu}(q)  =  i \int d^4 x e^{i q \cdot x} \langle 0 | T
\big{\{} j_\mu(x) j^\dagger_\nu(0) \big{\}} | 0 \rangle \; ,
\end{eqnarray}
where $j_\mu(x)$ is a current with the quantum numbers $J^{PC} = 1^{+-}$:
\begin{eqnarray}
j_\mu^{D_s \bar{D}_s^*} \!\!\!\!\!&=&\!\!\!\!\! \frac{i}{\sqrt{2}}\big[(\bar{s}_a \gamma_5 c_a)(\bar{c}_b
\gamma_\mu s_b) + (\bar{s}_a \gamma_\mu c_a)(\bar{c}_b \gamma_5 s_b)\big], \label{eq2} \\
j_\mu^{D_s^* \bar{D}_s^*} \!\!\!\!\!&=&\!\!\!\!\! \frac{i}{\sqrt{2}}\big[(\bar{s}_a \gamma^\alpha c_a)(\bar{c}_b
\sigma_{\alpha \mu} \gamma_5 s_b) -(\bar{s}_a \sigma_{\alpha \mu} \gamma_5 c_a)(\bar{c}_b \gamma^\alpha s_b)\big] ,
\end{eqnarray}
for respectively $D_s \bar{D}_s^*$ and $D_s^* \bar{D}_s^*$.
Here $a$, $b$ are color indices, and these currents keep closer relations respectively with $Z_c(3900)$ and $Z_c(4025)$ \cite{Cui:2013yva, Chen:2013omd}.

As $J_\mu(x)$ is not a conserved current, the two-point correlation function has two independent Lorentz structures \cite{Matheus:2006xi}:
\begin{eqnarray}
\Pi_{\mu\nu}(q) = - \left(g_{\mu\nu} -\frac{q_\mu q_\nu}{q^2}\right) \Pi_1(q^2)
+ \frac{q_\mu q_\nu}{q^2} \Pi_0(q^2) \; ,
\end{eqnarray}
where the invariant functions $\Pi_0(q^2)$ and $\Pi_1(q^2)$ are respectively related to the spin-0 and spin-1 mesons. In order to study the $1^{+-}$ molecular state, $\Pi_1(q^2)$ was utilized.

The fundamental assumption of the QCD Sum Rules is the principle of quark-hadron duality. Accordingly, on the one hand, the correlation function $\Pi_1(q^2)$ is obtained at the hadron level where the mass and coupling constant of the hadron are introduced. It may be calculated at the quark-gluon level, in which the Operator Product Expansion (OPE) is employed.

On the hadron side, after separating out the ground state contributions from the pole terms, the correlation function $\Pi_1(q^2)$ is obtained as a dispersion integral over a physical regime,
\begin{eqnarray}
\Pi_1(q^2) \!\!\!\!\!& = &\!\!\!\!\!\frac{\lambda_{H}^2}{M_{H}^2 -
q^2} + \frac{1}{\pi} \int_{s_0}^\infty d s \frac{\rho^h_H(s)}{s - q^2}
\;, \label{hadron}
\end{eqnarray}
in which $M_{H}$ is the mass of $D_s \bar{D}_s^*$ or $D_s^* \bar{D}_s^*$ molecular
state with $J^{PC} = 1^{+-}$, and $\rho^h_H(s)$
is the spectral density which contains contributions from the
higher excited states and the continuum states, $s_0$ is the
threshold of the higher excited states and continuum states, and the coupling constant $\lambda_{H}$ is defined by $\langle 0 | j_\mu |Z_{cs}\rangle = \lambda_{H} \epsilon_\mu$, where $Z_{cs}$ is the lowset lying $1^{+-}$ $D_s \bar{D}_s^*$ or $D_s^* \bar{D}_s^*$ molecular state.

On the quark-gluon side, the correlation function $\Pi_1(q^2)$ can be expressed as a dispersion relation:
\begin{eqnarray}
\Pi_{1}^{OPE}(q^2) &=& \int_{(2m_c + 2m_s)^2}^{\infty} d s
\frac{\rho^{OPE}(s)}{s - q^2} + \Pi_1^{\langle g_s^3 G^3 \rangle }(q^2) \nonumber \\
&+& \Pi_1^{\langle
\bar{s} s \rangle^2} (q^2) + \Pi_1^{\langle \bar{s} s \rangle \langle \bar{s} G s \rangle}(q^2) \; ,
\end{eqnarray}
in which $\rho^{OPE}(s) = \text{Im} [\Pi_1^{OPE}(s)] / \pi$ and
\begin{eqnarray}
\rho^{OPE}(s) &\!\!\!=\!\!\!&  \rho^{pert}(s) + \rho^{\langle \bar{s} s
\rangle}(s) + \rho^{\langle g_s^2 G^2 \rangle}(s) + \rho^{\langle \bar{s} G s \rangle}(s) \nonumber \\
&\!\!\!+\!\!\!& \rho^{\langle \bar{s} s
\rangle^2}(s) + \rho^{\langle g_s^3 G^3 \rangle}(s) + \rho^{\langle \bar{s} s \rangle \langle \bar{s} G s \rangle }(s)\; .
\end{eqnarray}
Here $\Pi_1^{\langle g_s^3 G^3 \rangle }(q^2)$, $\Pi_1^{\langle
\bar{s} s \rangle^2} (q^2)$ and $\Pi_1^{\langle \bar{s} s \rangle \langle \bar{s} G s \rangle}(q^2)$ denote those contributions of the correlation function which have no imaginary parts but have nontrivial values under the Borel transform. It should be mentioned that in principle the four-gluon operator, the $\langle g_s^2 G^2 \rangle^2$, also belongs to the dimension-eight condensate, however, in practice we find it is only 1\% of the mixed condensate $\langle \bar{s} s \rangle \langle \bar{s} G s \rangle $ in magnitude, and hence the four gluon condensate is neglected in the evaluation of this work.

After making a Borel transform of the quark-gluon side:
\begin{eqnarray}
\Pi_{1}^{OPE}(M_B^2) &\!\!\!\!=\!\!\!\!& \int_{(2m_c + 2m_s)^2}^{\infty} d s
\rho^{OPE}(s)e^{- s / M_B^2}  \nonumber \\
&\!\!\!\! + \!\!\!\!&  \Pi_1^{\langle g_s^3 G^3 \rangle }(M_B^2)  + \Pi_1^{\langle
\bar{s} s \rangle^2} (M_B^2) \nonumber \\
&\!\!\!\! + \!\!\!\!& \Pi_1^{\langle \bar{s} s \rangle \langle \bar{s} G s \rangle}
(M_B^2) \; . \label{spectral-density}
\end{eqnarray}

To consider the effects induced by the mass of the strange quark, terms which are linear
in the strange quark mass $m_s$ are utilized in the following calculations. For both the $D_s \bar{D}_s^*$ and $D_s^* \bar{D}_s^*$ molecular states, we put the concrete forms of spectral densities in eq.(\ref{spectral-density}) into the Appendix.

Performing the Borel transform on the hadron side (eq.(\ref{hadron})) and matching it to eq.(\ref{spectral-density}), the resultant sum rule for the mass of the hidden strange molecular state H with $1^{+-}$ is:
\begin{eqnarray}
M_{H}(s_0, M_B^2) = \sqrt{- \frac{R_1(s_0, M_B^2)}{R_0(s_0,
M_B^2)}} \; ,
\end{eqnarray}
where H represents the $D_s \bar{D}_s^*$ or $D_s^* \bar{D}_s^*$ molecular state and
\begin{eqnarray}
R_0(s_0, M_B^2) & \!\!\!\! = \!\!\!\! & \int_{(2m_c + 2m_s)^2}^{s_0} d s \; \rho^{OPE}(s) e^{-
s / M_B^2} + \Pi_1^{\langle g_s^3 G^3 \rangle }(M_B^2) \nonumber \\
&\!\!\!\! + \!\!\!\!&  \Pi_1^{\langle \bar{s} s \rangle^2} (M_B^2) + \Pi_1^{\langle \bar{s} s
\rangle \langle  \bar{s} G s \rangle}(M_B^2) \; , \label{momentum} \\
R_1(s_0, M_B^2) & \!\!\!\! = \!\!\!\! &
\frac{\partial}{\partial{M_B^{-2}}}{R_0(s_0, M_B^2)} \; .
\end{eqnarray}

\section{Numerical Results}

In the numerical calculation, the values of the condensates and the quark masses are used as \cite{Matheus:2006xi, Narison:2002pw, Nielsen:2009uh}:
$m_s = (0.13 \pm 0.03) \; \text{GeV}$, $m_c (m_c) = (1.23 \pm 0.05) \; \text{GeV}$, $m_b (m_b) = (4.24 \pm 0.06 ) \; \text{GeV}$, $\langle \bar{q} q \rangle = - (0.23 \pm 0.03)^3 \; \text{GeV}^3$, $\langle \bar{s} s \rangle = (0.8 \pm 0.2) \langle \bar{q} q \rangle$, $\langle g_s^2 G^2 \rangle = 0.88 \; \text{GeV}^4$, $\langle \bar{s} g_s \sigma \cdot G s
\rangle = m_0^2 \langle \bar{s} s \rangle$, $\langle g_s^3 G^3
\rangle = 0.045 \; \text{GeV}^6$, and $m_0^2 = 0.8 \; \text{GeV}^2$. Here, the strange quark mass is the current quark mass in a mass-independent substraction scheme such as $\overline{MS}$ at the scale $\mu = m_c$, whereas the charm and bottom quark masses are the running masses in the $\overline{MS}$ scheme. To provide greater clarity of choosing these parameters, the primary sources for determining these parameters can be found in refs.~\cite{Chen:2010ze, Eidemuller:2000rc, Jamin:2001zr}.

To determinate the Borel parameter $M_B^2$ and the threshold parameter $s_0$, we used the following limit constraints, which is the standard procedure for QCD Sum Rules analysis. The sum rule parameter $\tau = 1/M_B^2$ is utilized in our analysis. In the QCD Sum Rules, for choosing the threshold $s_0$ and the parameter $\tau$, there are two criteria \cite{Shifman,
Reinders:1984sr, P.Col}. First, the convergence of the OPE
is retained. Therefore, in order to determine their convergence,
one needs to compare the relative contributions of each term to the total contributions of the OPE side. The second criterion to constrain the $\tau$ is that the pole contribution (PC), defined as the pole contribution divided by the total contribution (pole plus continuum), is larger than the continuum contribution. In order to safely eliminate the contributions of the higher excited and continuum states, the PC is generally greater than
$50\%$ \cite{P.Col, Matheus:2006xi}.

To find a proper value for $\sqrt{s_0}$, we carry out a similar analysis as in ref.~\cite{Finazzo:2011he}. Since the continuum threshold is connected to the mass of the studied state by the relation $\sqrt{s_0} \sim (M_H + \delta) \text{GeV}$, where $\delta$ is about 0.6 GeV, various $\sqrt{s_0}$ satisfying this constraint are taken into account. Among these values, one needs then to find out the proper one which has an optimal window for Borel parameter $M_B^2$. That is, within this window, the physical quantity, here the molecular mass $M_H$, is independent of the Borel parameter $M_B^2$ as much as possible. Through the above procedure one obtains the central value of $\sqrt{s_0}$. However, in practice, in the QCD Sum Rules calculation, it is normally acceptable to vary the $\sqrt{s_0}$ by $0.1 \text{GeV}$~\cite{Finazzo:2011he}, which gives the lower and upper bounds and hence the uncertainties of $\sqrt{s_0}$.

\begin{figure}[htb!]
\includegraphics[width=8.5cm]{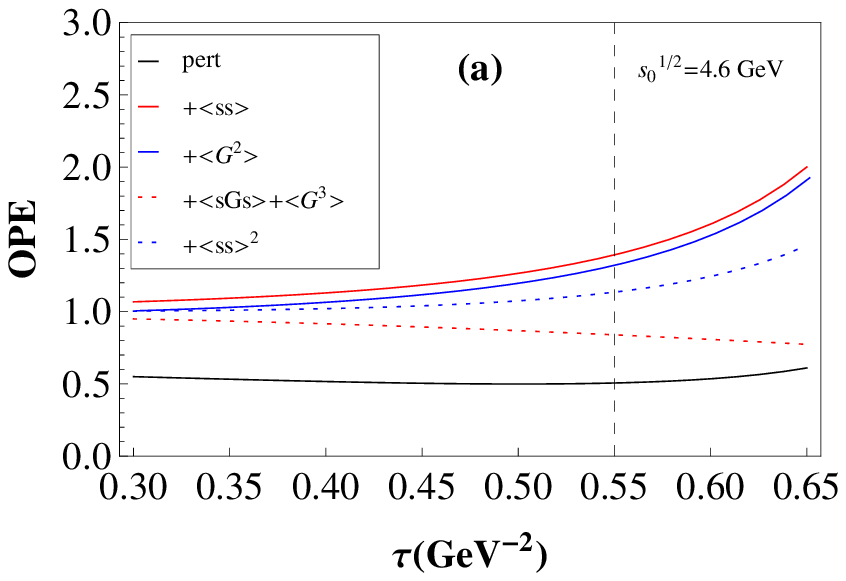}
\includegraphics[width=8.5cm]{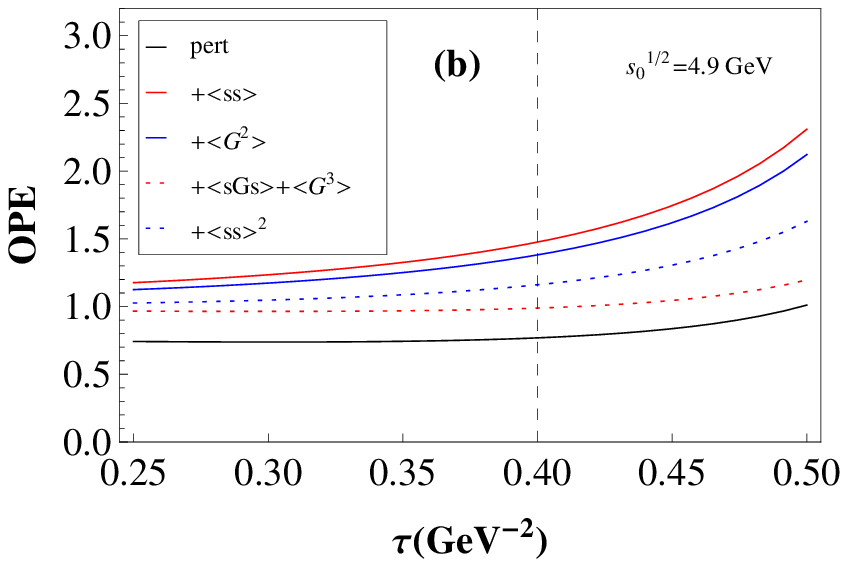}
\caption{(\textbf{a}) OPE convergence in the region $0.30 \leq \tau \leq 0.65 \; \text{GeV}^{-2}$ for the $D_s \bar{D}_s^*$ molecular state with $\sqrt{s_0} = 4.6 \; \text{GeV}$, where $\tau = 1/M_B^2$.
(\textbf{b}) OPE convergence in the region $0.25 \leq \tau \leq 0.50 \; \text{GeV}^{-2}$ for the $D_s^* \bar{D}_s^*$ molecular state with $\sqrt{s_0} = 4.9 \; \text{GeV}$. The black line denotes the fraction of perturbative contribution, and each subsequent line denotes the addition of one extra condensate, {\it i.e.}, $+ \langle \bar{s} s \rangle$ (red line), $+ \langle g_s^2 G^2 \rangle$ (blue line), $+ \langle g_s \bar{s} \sigma \cdot G s \rangle$ + $\langle g_s^3 G^3 \rangle$ (red dotted line), $+ \langle \bar{s} s \rangle^2$ (blue dotted line). The vertical lines respectively denote the upper limit constraints of the $\tau$ for $D_s \bar{D}_s^*$ and $D_s^* \bar{D}_s^*$. The curve including the dimension-eight contribution is defined as :($pert + \langle \bar{s} s \rangle + \langle g_s^2 G^2 \rangle + \langle g_s \bar{s} \sigma \cdot G s \rangle + \langle g_s^3 G^3 \rangle + \langle \bar{s} s \rangle^2 + \langle \bar{s} s \rangle \langle g_s \bar{s} \sigma \cdot G s\rangle$)/total. Hence the curve including dimension eight operator is just a straight line, and we do not show it here.}  \label{convergence}
\end{figure}

The OPE convergences are illustrated in fig.\ref{convergence} respectively for $D_s \bar{D}_s^*$ and $D_s^* \bar{D}_s^*$. Complying with the first criterion, the upper limit constraints of $\tau$ are $\tau \leq 0.55 \; \text{GeV}^{-2}$ and $\tau \leq 0.40 \; \text{GeV}^{-2}$ with $\sqrt{s_0} = 4.6 \, \text{GeV}$ and $\sqrt{s_0} = 4.9 \, \text{GeV}$.

\begin{figure}[htb!]
\includegraphics[width=8.5cm]{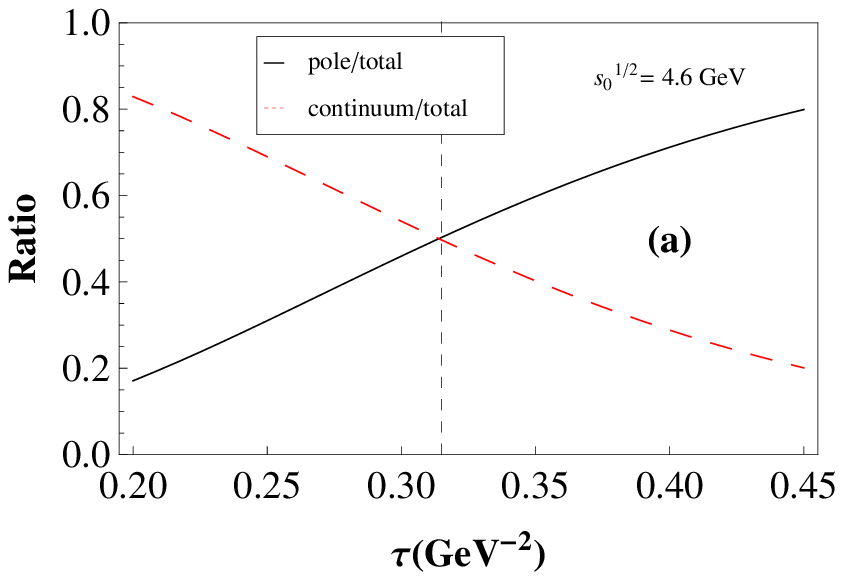}
\includegraphics[width=8.5cm]{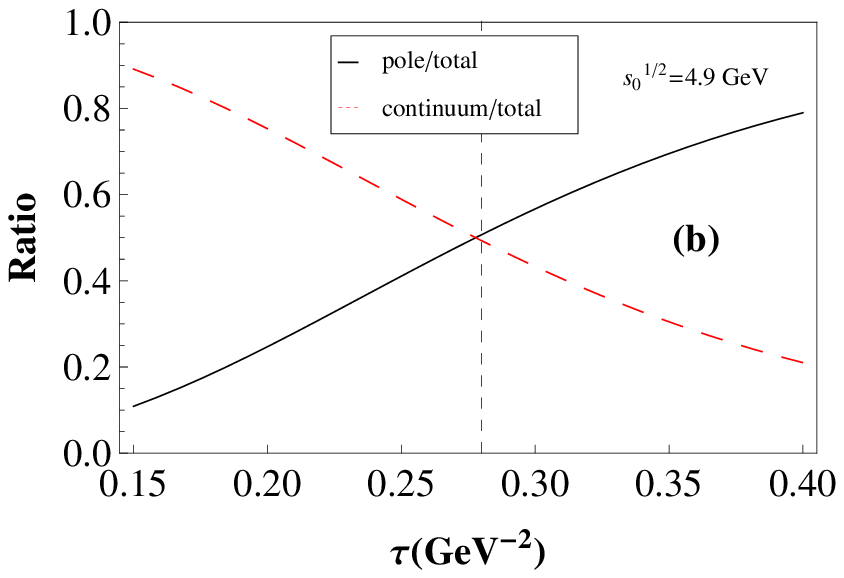}
\caption{(\textbf{a}) The relative pole contribution of $D_s \bar{D}_s^*$ state with $\sqrt{s_0} = 4.6 \; \text{GeV}$. (\textbf{b}) The relative pole contribution of $D_s^* \bar{D}_s^*$ state with $\sqrt{s_0} = 4.9 \; \text{GeV}$. The black line represents the relative contribution, whereas the red dashed line corresponds to the continuum contribution. The vertical lines respectively denote the lower limit constraints of the $\tau$ for $D_s \bar{D}_s^*$ and $D_s^* \bar{D}_s^*$.} \label{pole}
\end{figure}

The result of the PC is shown in fig.\ref{pole}, which indicates
the lower limit constraint of $\tau$. Noting that the lower
limit constraint of $\tau$ depends on the threshold value
$s_0$, for different $s_0$, there are different lower limits
of $\tau$. To determine the value of $s_0$, an analysis
similar to ref.\cite{Matheus:2006xi} was carried out.

\begin{figure}[htb!]
\includegraphics[width=8.5cm]{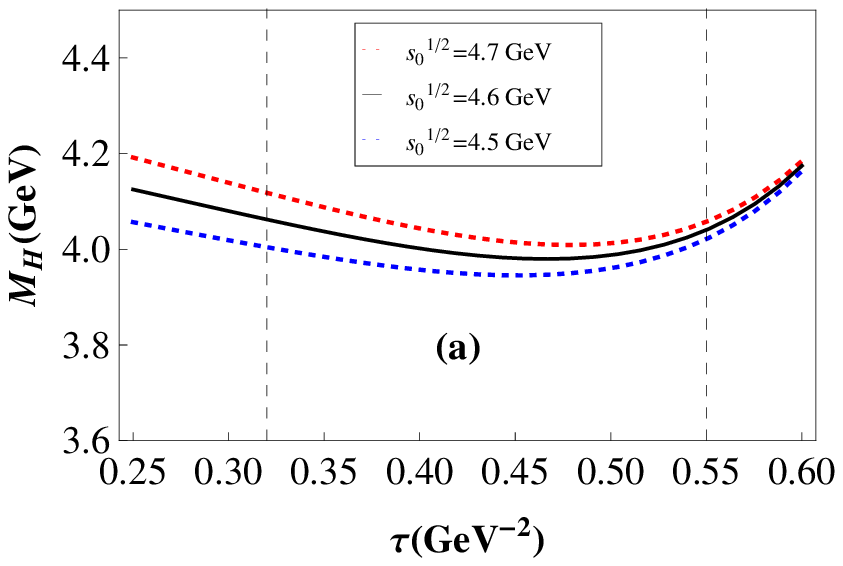}
\includegraphics[width=8.5cm]{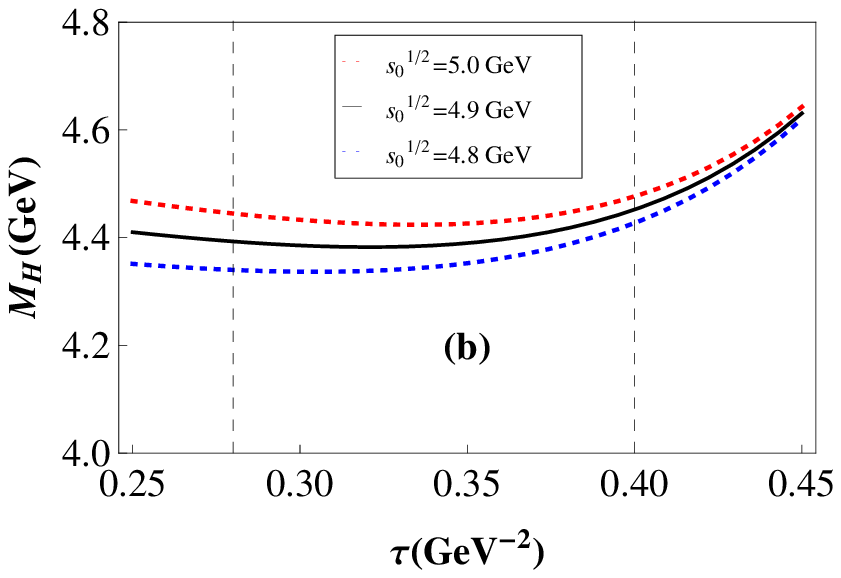}
\caption{(\textbf{a}) The mass of the $D_s \bar{D}_s^*$ state as a function of the sum rule parameter $\tau$, for different values of $\sqrt{s_0}$. (\textbf{b}) The mass of the $D_s^* \bar{D}_s^*$ state as a function of the sum rule paramter $\tau$, for different values of $\sqrt{s_0}$. The two vertical lines indicate the upper and lower limits of a valid Borel window.} \label{mass}
\end{figure}

The dependence of $M_H$ on the parameter $\tau$ in fig.\ref{mass} with various $\sqrt{s_0}$ is drawn where the two
vertical lines indicate the upper and lower limits of a valid Borel window for the central value of $\sqrt{s_0}$. For complete analysis, the Borel windows with various $\sqrt{s_0}$ are listed in table.\ref{window}.

\begin{table*}[htb!]
\caption{The lower and upper limit constraints of the Borel parameter $\tau$ for the $D_s \bar{D}_s^*$ and $D_s^* \bar{D}_s^*$ molecular states with different values of $\sqrt{s_0}$.}
\begin{center}
\renewcommand\arraystretch{1.1}
\begin{tabular}{|c|c|c|c|c|c|}
\hline \multicolumn{3}{|c|}  {$D_s \bar{D}_s^*$}  &  \multicolumn{3}{|c|}
{$D_s^* \bar{D}_s^*$}  \\
\hline $\sqrt{s_0} \, (\text{\footnotesize{GeV}})$ & $\tau_{min} \,
(\text{\footnotesize{GeV}}^{-2})$  & $\tau_{max} \, (\text{\footnotesize{GeV}}^{-2})$ & $\sqrt{s_0} \,
(\text{\footnotesize{GeV}})$  & $\tau_{min} \, (\text{\footnotesize{GeV}}^{-2})$ & $\tau_{max} \, (\text{\footnotesize{GeV}}^{-2})$ \\
\hline 4.7 & 0.30 & 0.58 & 5.0 & 0.26 & 0.42 \\
\hline 4.6 & 0.32 & 0.55 & 4.9 & 0.28 & 0.40\\
\hline 4.5 & 0.33 & 0.53 & 4.8 & 0.30 & 0.38 \\
\hline
\end{tabular}
\end{center}
\label{window}
\end{table*}

Eventually, the mass of the $D_s \bar{D}_s^*$ molecular state was determined to be:
\begin{eqnarray}
M_H^{D_s \bar{D}_s^*} = (3.98 \pm 0.15) \, \text{GeV} \; ,
\end{eqnarray}
where the mass with the optimal stability was extracted with errors stemming from the uncertainties of the quark mass, the condensates, the Borel parameter and the threshold parameter $\sqrt{s_0}$.

The central value for the mass of the $D_s \bar{D}_s^*$ molecular state is below that of the meson-meson threshold $E_{th}[D_s \bar{D}_s^*] \simeq 4.08 \, \text{GeV}$ by 100 MeV, where the notation $E_{th}[M_1 \, M_2]$ represents the corresponding energy of the sum of the masses of the $M_1$ and $M_2$ mesons. Therefore the $D_s \bar{D}_s^* (1^{+-})$ molecular state forms a bound state, which predicts a hidden strange charmonium-like state around $3.98 \, \text{GeV}$.

For the $D_s^* \bar{D}_s^*$ molecular state:
\begin{eqnarray}
M_H^{D_s^* \bar{D}_s^*} = (4.38 \pm 0.16) \, \text{GeV} \; .
\end{eqnarray}

The central value for the mass of the $D_s^* \bar{D}_s^*$ molecular state is above that of the meson-meson threshold $E_{th}[D_s^* \bar{D}_s^*] \simeq 4.22 \, \text{GeV}$
by 160 MeV. Therefore, the $D_s^* \bar{D}_s^* (1^{+-})$  molecular state forms a resonance, which predicts another hidden strange charmonium-like state around $4.38 \, \text{GeV}$.

It is important to compare our results with those obtained in refs.~\cite{Cui:2013yva, Chen:2013omd}. The strange quark effect is not huge, and since we take the similar constraints as in ref.~\cite{Cui:2013yva}, our result on $D \bar{D}^*$ agrees with ref.~\cite{Cui:2013yva} in the massless limit for strange quark. But in $D_s^*\bar{D}_s^*$ case, even in the massless limit our result is 0.25 GeV higher than that in ref.~\cite{Chen:2013omd} on $D^*\bar{D}^*$. This is due to the different constraint criteria between this work and \cite{Chen:2013omd}.

\begin{table*}[htb!]
\caption{The effect of a on-vanishing strange quark mass in the molecular states $D_s\bar{D}_s^*$ and $D_s^* \bar{D}_s^*$.}
\begin{center}
\renewcommand\arraystretch{1.1}
\begin{tabular}{|c|c|c|c|c|}
\hline  & $\sqrt{s_0}$(GeV) & $\tau_{min}$(GeV$^{-2}$) & $\tau_{max}$(GeV$^{-2}$)  & mass (GeV)  \\
\hline $D_s\bar{D}_s^*$ & 4.6  & 0.32 & 0.55 & $3.98\pm0.15$  \\
\hline $D\bar{D}^*$ & 4.4 & 0.34 & 0.51 & $3.88\pm0.18$ \\
\hline $D_s^*\bar{D}_s^*$ & 4.9 & 0.28 & 0.40 & $4.38\pm0.16$  \\
\hline $D^*\bar{D}^*$ & 4.7  & 0.31  & 0.40 & $4.29\pm0.19$ \\
\hline
\end{tabular}
\end{center}
\label{comparison}
\end{table*}

Ref.~\cite{Chen:2010ze} systematically evaluated the masses of the tetraquark systems with various quantum numbers and hidden-charm. The results for the hidden-charm and strange tetra- quark state with $J^{PC}=1^{+-}$ are in the region of 3.92-4.34 GeV. Comparing to our results, the masses of the molecular states in our calculation lie in 3.83-4.54 GeV, which covers the tetraquark mass region. Since considerable errors exist, so far we think it is hard to distinguish the underlying quark configurations of the tetraquark states and molecular states.

\section{Summary}

This work used the QCD Sum Rules to study the masses of the $D_s \bar{D}_s^*$  and $D_s^* \bar{D}_s^*$ molecular states with quantum numbers $J^{PC} = 1^{+-}$. In our calculation, interpolating currents with definite charge parity were employed, and the contributions up to dimension eight in the Operator Product Expansion (OPE) were taken into account. The numerical results were respectively $(3.98 \pm 0.15) \, \text{GeV}$ and $(4.38 \pm 0.16) \, \text{GeV}$ for the $D_s \bar{D}_s^*$ and $D_s^* \bar{D}_s^*$ molecular states.

The central value of $D_s \bar{D}_s^*$ molecular state was below the corresponding meson-meson threshold $E_{th}[D_s \bar{D}_s^*] \simeq 4.08 \, \text{GeV}$  about 100 MeV, which means that such molecular state would be tightly bound. The reason for the relative large binding energy is that the current in eq.(\ref{eq2}) is local. Hence they do not represent a object with two mesons separated in space, but rather compact one with two singlet quark-antiquark pairs. The central value of $D_s^* \bar{D}_s^*$ molecular state was above that of its corresponding meson-meson threshold $E_{th}[D_s^* \bar{D}_s^*] \simeq 4.22 \, \text{GeV}$.
Therefore, the $D_s^* \bar{D}_s^*$ molecular state with $1^{+-}$ may form a resonance.

In conclusion, considering the uncertainties, our results indicated that two hidden strange charmonium-like states, which may be probed in future experiments, may exist in the energy ranges of $3.83 \sim 4.13 \; \text{GeV}$ and $4.22 \sim 4.54 \; \text{GeV}$.

To ascertain the hidden strange charmonium-like states through their decays, the following decay modes may be measured: $e^+ e^- \to (\eta_c + \omega) + \eta$, $e^+ e^- \to (J/\psi + \eta) + \eta$ and $e^+ e^- \to (J/\psi + f_0(500)) + \eta$ for $D_s \bar{D}_s^*$ molecular state, and $e^+ e^- \to (\eta_c + \phi) + \eta$, $e^+ e^- \to (J/\psi + \eta^\prime) + \eta$, $e^+ e^- \to (J/\psi + f_0(980)) + \eta$, $e^+ e^- \to (h_c + \eta) + \eta$, $e^+ e^- \to (h_c + f_0(500)) + \eta$ and $e^+ e^- \to (D_s^* + \bar{D}_s^*) + \eta$ for $D_s^* \bar{D}_s^*$ molecular state.

We suggest future experiments to search for these hidden strange
charmonium-like resonances. The BESIII, Belle and BaBar and forthcoming BelleII and SuperB will facilitate searching for such hidden strange charmonium-like resonances.

\vspace{.7cm} {\bf Acknowledgments} \vspace{.3cm}

We thank C.-Z.Yuan for his comments on this work.
This work was supported in part by the National Natural Science Foundation of China(NSFC) under the grants 10935012, 11175249, 11121092, and 11375200.

\appendix{\bf\Large Appendix}

For the $D_s \bar{D}_s^*$ molecular state, the concrete forms of spectral densities in eq.(\ref{spectral-density}) are obtained as:
\begin{eqnarray*}
\rho^{pert}_{D_s \bar{D}_s^*}(s) \!\!\! &=& \!\!\! \frac{3}{2^{12} \pi^6}
\int_{\alpha_{min}}^{\alpha_{max}} \frac{d
\alpha}{\alpha^3} \int_{\beta_{min}}^{1 - \alpha} \frac{d
\beta}{\beta^3} {\cal F}^3 (1 - \alpha - \beta) \bigg[(1 + \alpha + \beta) {\cal F} \nonumber \\
\!\!\! &-& \!\!\! 2(\alpha + \beta)(3 + \alpha + \beta)\bigg] \; , \\
\rho^{\langle \bar{s} s \rangle}_{D_s \bar{D}_s^*}(s) \!\!\! &=& \!\!\!
\frac{3 \langle \bar{s} s \rangle}{2^8 \pi^4} \int_{\alpha_{min}}^{\alpha_{max}} d
\alpha \bigg{\{} \frac{2 m_s {\cal H}^2}{(1 - \alpha) \alpha} - \int_{\beta_{min}}^{1 - \alpha}
\frac{d \beta}{\alpha \beta} {\cal F} \bigg[ [m_c (1 + \alpha + \beta) \frac{(\alpha + \beta)}
{\alpha \beta} \nonumber \\
\!\!\! &+& \!\!\! 2 m_s \alpha \beta] {\cal F} + 8 m_s m_c^2 \bigg] \bigg{\}}\;,
\end{eqnarray*}
\begin{eqnarray*}
\rho^{\langle g_s^2 G^2 \rangle}_{D_s \bar{D}_s^*}(s) \!\!\! &=& \!\!\! -
\frac{\langle g_s^2 G^2 \rangle}{2^{13} \pi^6} \int_{\alpha_{min}}^{\alpha_{max}} \frac{d
\alpha}{\alpha^3} \int_{\beta_{min}}^{1 - \alpha} \frac{d
\beta}{\beta^3} \bigg[ 2 \alpha \beta (\alpha + \beta) (1 - 2\alpha - 2\beta) {\cal F}^2 \nonumber \\
\!\!\! &+& \!\!\! \big[4m_s m_c \alpha^2 \beta^2 (1 + \alpha + \beta) + 3m_s m_c
(\alpha^3 + \beta^3)(1 - \alpha - \beta)(3 + \alpha + \beta) \nonumber \\
\!\!\! &-& \!\!\! 2 m_c^2 (\alpha + \beta)(1-(\alpha + \beta)^2)(\alpha^2 -2\alpha
\beta + \beta^2)\big] {\cal F} \nonumber \\
\!\!\! &+& \!\!\! m_s m_c^3 (\alpha + \beta)^2 (1 - \alpha - \beta)(3 + \alpha + \beta)
(\alpha^2 - \alpha \beta + \beta^2)\bigg] \; , \\
\rho^{\langle \bar{s} G s \rangle}_{D_s \bar{D}_s^*}(s) \!\!\! &=& \!\!\!
\frac{\langle g_s \bar{s} \sigma \cdot G s \rangle}{2^7 \pi^4}
\int_{\alpha_{min}}^{\alpha_{max}} d \alpha \bigg{\{} 2 m_s ({\cal H} + m_c^2) -
\frac{1}{\alpha}(3 m_c + m_s) {\cal H} \nonumber \\
\!\!\! &+& \!\!\! \int_{\beta_{min}}^{1 - \alpha} d \beta \bigg[\frac{3 m_c}{2}
\frac{{\cal F}}{\alpha} + \frac{1}{\alpha^2}\big[(2 m_s \alpha +3 m_c (\alpha +
\beta)){\cal F} - 3 m_s m_c^2 \alpha \big]\bigg]\bigg{\}} \; , \\
\rho^{\langle g_s^3 G^3 \rangle}_{D_s \bar{D}_s^*}(s) \!\!\! &=& \!\!\!
\frac{\langle g_s^3 G^3 \rangle}{2^{14} \pi^6} \int_{\alpha_{min}}^{\alpha_{max}}
\frac{d \alpha}{\alpha^3} \int_{\beta_{\min}}^{1 - \alpha} d \beta \bigg[(1 -
(\alpha + \beta)^2) (2 {\cal F} + 4m_c^2 \beta ) \nonumber \\
\!\!\! &-& \!\!\! m_s m_c (1 - \alpha -\beta)(3 + \alpha + \beta)(\alpha + 6\beta)\bigg]\;, \\
\rho^{\langle \bar{s} s \rangle^2}_{D_s \bar{D}_s^*}(s) \!\!\! &=& \!\!\!
\frac{\langle \bar{s} s \rangle^2}{2^8 \pi^2} \bigg[4 m_c^2 - 3 m_s m_c \bigg]
\sqrt{1 - \frac{4 m_c^2}{s}} \; ,\\
\Pi^{\langle g_s^3 G^3\rangle}_{D_s \bar{D}_s^*} (M_B^2) \!\!\! &=&
\!\!\! \frac{m_s m_c^3 \langle g_s^3 G^3 \rangle}{2^{13} \pi^6} \int_{0}^{1}
\frac{d \alpha}{\alpha^4} \int_{0}^{1 - \alpha} d \beta (\alpha + \beta)(1 -
\alpha - \beta) \nonumber \\
\!\!\! &\times& \!\!\! (3 + \alpha + \beta) \text{exp} \bigg[ - \frac{m_c^2
(\alpha + \beta)}{M_B^2 \alpha \beta}\bigg] \; , \\
\Pi^{\langle \bar{s} s \rangle^2}_{D_s \bar{D}_s^*} (M_B^2) \!\!\! &=& \!\!\! -
\frac{m_s m_c^3 \langle \bar{s} s \rangle^2}{2^4 \pi^2} \int_{0}^{1}
\frac{d \alpha}{\alpha} \text{exp} \bigg[ - \frac{m_c^2}{M_B^2 (1 - \alpha) \alpha}\bigg] \; ,
\end{eqnarray*}
and the contribution from the dimension-eight condensate is obtained as:
\begin{eqnarray*}
\Pi^{\langle \bar{s} s \rangle \langle \bar{s} G s \rangle}_{D_s \bar{D}_s^*} (M_B^2)
\!\!\! &=& \!\!\! \frac{m_c \langle \bar{s} s \rangle \langle \bar{s} \sigma \cdot G s
\rangle}{3 \times 2^8 \pi^2} \int_{0}^{1} \frac{d \alpha}{(1 - \alpha) \alpha}
\bigg[ 12 m_c (1 - 2 (1 - \alpha)\alpha ) \nonumber \\
\!\!\! &+& \!\!\! 11 m_s (1 - \alpha) \alpha + \frac{m_c^2}{(1 - \alpha) \alpha}
\big[- 24 m_c (1 - \alpha) \alpha \nonumber \\
\!\!\! &+& \!\!\! m_s (19 (1-\alpha) \alpha - 6)\big] \bigg] \text{exp}\bigg[ -
\frac{m_c^2}{M_B^2 (1 - \alpha) \alpha}\bigg] \; .
\end{eqnarray*}
Here, $M_B$ is the Borel parameter introduced by the Borel
transformation, ${\cal F} (\alpha, \beta, s) = (\alpha + \beta)
m_c^2 - \alpha \beta s$, ${\cal H} (\alpha, s) = m_c^2 - \alpha (1 - \alpha) s$ and the integration limits are given by $\alpha_{min} = (1 - \sqrt{1 - 4 m_c^2/s}) / 2$, $\alpha_{max} = (1 + \sqrt{1 - 4 m_c^2 / s}) / 2$ and $\beta_{min} = \alpha m_c^2 /(s \alpha - m_c^2)$.

For the $D_s^* \bar{D}_s^*$ molecular state, the concrete forms of spectral densities in eq.(\ref{spectral-density}) are obtained as:
\begin{eqnarray*}
\rho^{pert}_{D_s^* \bar{D}_s^*}(s) \!\!\! &=& \!\!\! \frac{1}{2^{11} \pi^6}
\int_{\alpha_{min}}^{\alpha^{max}} \frac{d \alpha}{\alpha^3}
\int_{\beta_{min}}^{1 - \alpha} \frac{d \beta}{\beta^3} \bigg[ \frac{9}{2}
(1 - (\alpha + \beta)^2) {\cal F} - 2m_c^2 (1 - \alpha - \beta)^2
(5 + \alpha + \beta) \nonumber \\
\!\!\! &-& \!\!\! 9m_s m_c (\alpha + \beta)(1 - \alpha - \beta)
(3 + \alpha + \beta)  \bigg] {\cal F}^3 \; , \\
\rho^{\langle \bar{s} s \rangle}_{D_s^* \bar{D}_s^*}(s) \!\!\! &=
& \!\!\! -\frac{3 \langle \bar{s} s \rangle}{2^8 \pi^4}
\int_{\alpha_{min}}^{\alpha_{max}} d \alpha \bigg{\{} \frac{2m_s {\cal H}^2}
{(1 - \alpha)\alpha} + \int_{\beta_{min}}^{1 - \alpha} d \beta
\bigg[ \big[\frac{3m_c (\alpha + \beta)(1 + \alpha + \beta)}{\alpha^2 \beta^2}
+ \frac{2 m_s}{\alpha \beta}\big] {\cal F} \nonumber \\
\!\!\! &-& \!\!\! \frac{4m_s m_c^2 (5 - \alpha - \beta)}
{\alpha \beta}\bigg] {\cal F} \bigg{\}} \; , \\
\rho^{\langle g_s^2 G^2 \rangle}_{D_s^* \bar{D}_s^*}(s) \!\!\! &=
& \!\!\! \frac{\langle g_s^2 G^2 \rangle}{2^{13} \pi^6}
\int_{\alpha_{min}}^{\alpha_{max}} \frac{d \alpha}{\alpha^2}
\int_{\beta_{min}}^{1 - \alpha} \frac{d \beta}{\beta^2}
\bigg[ 2(\alpha + \beta)(1 - 2\alpha - 2\beta) {\cal F}^2 \nonumber \\
\!\!\! &+& \!\!\! \frac{2m_c^2 (1 - \alpha - \beta)}{\alpha \beta}[4( \alpha^4 + \beta^4) + 6\alpha \beta (\alpha^2 + \beta^2) + 4\alpha^2 \beta^2 + 7(\alpha^3 + \beta^3) \nonumber \\
\!\!\! &+& \!\!\! 7\alpha \beta(\alpha + \beta) - 5(\alpha^2 + \beta^2)]{\cal F} + \frac{m_s m_c}{\alpha \beta}[9(\alpha^5 + \beta^5) + 22\alpha \beta(\alpha^3 + \beta^3) \nonumber \\
\!\!\! &+& \!\!\! 18 (\alpha^4 + \beta^4) + 25\alpha^2 \beta^2 (\alpha + \beta) + 22\alpha \beta(\alpha^2 + \beta^2) - 27 (\alpha^3 + \beta^3) \nonumber \\
\!\!\! &+& \!\!\! 12\alpha^2 \beta^2]{\cal F} - \frac{2m_c^4 }{3\alpha \beta}(\alpha^3 + \beta^3)(1 - \alpha - \beta)^2 (5 + \alpha + \beta) \nonumber \\
\!\!\! &-& \!\!\! \frac{3m_s m_c^3}{\alpha \beta}(1 - \alpha - \beta)(\alpha + \beta)(\alpha^3 + \beta^3)(3 + \alpha + \beta) \bigg] \; , \\
\rho^{\langle \bar{s} G s \rangle}_{D_s^* \bar{D}_s^*}
(s) \!\!\! &=& \!\!\! \frac{\langle g_s \bar{s} \sigma
\cdot G s \rangle}{2^8 \pi^4} \int_{\alpha_{min}}^{\alpha_{max}}
d \alpha \bigg{\{} 20 m_s m_c^2 - \frac{m_s {\cal H}}{\alpha(1 - \alpha)} [1 + 18 \alpha (1 - \alpha)] -
\frac{9 m_c {\cal H}}{(1 - \alpha)\alpha} \nonumber \\
\!\!\! &+& \!\!\! \int_{\beta_{min}}^{1 - \alpha} d
\beta \bigg[\frac{m_c}{2\alpha^2 \beta^2} [2(\alpha^3 + \beta^3)
+ 7\alpha \beta (\alpha + \beta) - 4 (\alpha^2 + \beta^2) -
4\alpha \beta] {\cal F} \nonumber \\
\!\!\! &-& \!\!\! \frac{m_s (\alpha + \beta)}{\alpha \beta}
{\cal F} - \frac{m_s m_c^2}{\alpha \beta}[(\alpha^2 + \beta^2)-2(\alpha + \beta)] \bigg]\bigg{\}} \; ,\\
\rho^{\langle g_s^3 G^3 \rangle}_{D_s^* \bar{D}_s^*}(s)
\!\!\! &=& \!\!\! \frac{\langle g_s^3 G^3 \rangle }{2^{13}
\pi^6} \int_{\alpha_{min}}^{\alpha_{max}} \frac{d \alpha}{\alpha^3}
\int_{\beta_{min}}^{1 - \alpha} \frac{d \beta}{\beta^3} \bigg[\frac{3{\cal F}}{2}[1 - (\alpha + \beta)^2](\alpha^3 + \beta^3) + m_c^2(1 - \alpha - \beta) \nonumber \\
\!\!\! &\times& \!\!\! [4(\alpha^5 + \beta^5) + 5\alpha \beta (\alpha^3 + \beta^3) + 7(\alpha^4 + \beta^4) + \alpha^2 \beta^2 (\alpha + \beta) + 4\alpha \beta (\alpha^2 + \beta^2) \nonumber \\
\!\!\! &-& \!\!\! 5(\alpha^3 + \beta^3)] - \frac{3m_s m_c}{4}(1 - \alpha - \beta)[6(\alpha^5 + \beta^5) + 7\alpha \beta (\alpha^3 + \beta^3) \nonumber \\
\!\!\! &+& \!\!\! 18(\alpha^4 + \beta^4) + \alpha^2 \beta^2 (\alpha + \beta) + 3\alpha \beta (\alpha^2 + \beta^2)]  \bigg] \; , \\
\rho^{\langle \bar{s} s \rangle^2}_{D_s^* \bar{D}_s^*}(s)
\!\!\! &=& \!\!\! -\frac{\langle \bar{s} s\rangle^2}{2^6 \pi^2}
\int_{\alpha_{min}}^{\alpha_{max}} d \alpha \big[20({\cal H} - m_c^2) + 9m_s m_c \big] \; ,
\end{eqnarray*}
\begin{eqnarray*}
\Pi^{\langle g_s^3 G^3 \rangle}_{D_s^* \bar{D}_s^*}(M_B^2)
\!\!\! &=& \!\!\! \frac{\langle g_s^3 G^2 \rangle}{3 \times 2^{13} \pi^6}
\int_{0}^{1} \frac{d \alpha}{\alpha^4} \int_0^{1 - \alpha} \frac{d
\beta}{\beta^4} \bigg[ m_c^4 (1 - \alpha - \beta)^2 [(\alpha^5 + \beta^5) + \alpha \beta(\alpha^3 + \beta^3) \nonumber \\
\!\!\! &+& \!\!\! 5(\alpha^4 + \beta^4)] + \frac{9m_s m_c^3}{2} (1 - \alpha - \beta) (\alpha + \beta) (3 + \alpha + \beta) (\alpha^4 + \beta^4) \bigg] \nonumber \\
\!\!\! &\times& \!\!\! \text{exp}\bigg[ -
\frac{m_c^2(\alpha + \beta)}{M_B^2 \alpha \beta} \bigg] \; , \\
\Pi^{\langle \bar{s} s \rangle^2}_{D_s^* \bar{D}_s^*}(M_B^2)
\!\!\! &=& \!\!\! -\frac{3m_s m_c^3 \langle \bar{s} s \rangle^2}
{2^5 \pi^2}\int_0^1 d \alpha \frac{1}{(1 - \alpha) \alpha}
\text{exp} \bigg[- \frac{m_c^2}{(1 - \alpha) \alpha M_B^2} \bigg] \; ,\\
\Pi^{\langle \bar{s} s \rangle \langle \bar{s}
G s \rangle}_{D_s^* \bar{D}_s^*}(M_B^2) \!\!\!
&=& \!\!\! \frac{\langle \bar{s} s \rangle \langle g_s \bar{s}
\sigma \cdot G s \rangle}{3 \times 2^8 \pi^2} \int_0^1 d \alpha
\bigg[- \frac{20m_c^2}{(1 - \alpha) \alpha}[1 + 12 \alpha (1 - \alpha)] \nonumber \\
\!\!\! &+& \!\!\! \frac{m_s m_c}{(1 - \alpha) \alpha}[8 + 45\alpha(1 - \alpha)] - \frac{120m_c^4}{(1 - \alpha)\alpha M_B^2} + \frac{m_s m_c^3}{(1 - \alpha)^2 \alpha^2 M_B^2} [10 \nonumber \\
\!\!\! &+& \!\!\ 57 \alpha (1 - \alpha)] + \frac{30m_s m_c^5}{(1 - \alpha)^2 \alpha^2 M_B^4} \bigg]
\text{exp} \bigg[ - \frac{m_c^2}{(1 - \alpha) \alpha M_B^2} \bigg] .
\end{eqnarray*}

\end{document}